\documentclass[twocolumn, prx, a4paper, 10pt, superscriptaddress, nopacs, floatfix]{revtex4-1}
\usepackage{siunitx}
\usepackage{graphicx}
\usepackage{xcolor}
\usepackage{physics}
\usepackage{nicefrac}
\usepackage{bm}
\usepackage{float}
\usepackage[utf8]{inputenc}
\usepackage[colorlinks, citecolor=blue, linkcolor=black]{hyperref}
\usepackage{ulem}
\usepackage{booktabs}
\usepackage{multirow}
\usepackage{xspace}

\newcommand{\VB}{V$_\text{B}^-$\xspace}

\begin{document}

\title{Spin-dependent photodynamics of boron-vacancy centers in hexagonal boron nitride}

\author{T. Clua-Provost}
\thanks{Contributed equally to this work.}
\author{Z. Mu}
\thanks{Contributed equally to this work.}
\author{A. Durand}
\author{C. Schrader}
\affiliation{Laboratoire Charles Coulomb, Universit\'e de Montpellier and CNRS, 34095 Montpellier, France}
\author{J. Happacher}
\author{J.~Bocquel}
\author{P. Maletinsky}
\affiliation{Department of Physics, University of Basel, Basel, Switzerland}
\author{J.~Frauni\'e}
\author{X.~Marie}
\author{C. Robert}
\affiliation{Universit\'e de Toulouse, INSA-CNRS-UPS, LPCNO, 135 Avenue Rangueil, 31077 Toulouse, France}
\author{G. Seine}
\affiliation{CEMES-CNRS and Universit\'e de Toulouse, 29 rue J. Marvig, 31055 Toulouse, France}
\author{E.~Janzen}
\author{J.~H.~Edgar}
\affiliation{Tim Taylor Department of Chemical Engineering, Kansas State University, Kansas 66506, USA}
\author{B. Gil}
\affiliation{Laboratoire Charles Coulomb, Universit\'e de Montpellier and CNRS, 34095 Montpellier, France}
\author{G. Cassabois}
\affiliation{Laboratoire Charles Coulomb, Universit\'e de Montpellier and CNRS, 34095 Montpellier, France}
\affiliation{Institut Universitaire de France, 75231 Paris, France}
\author{V. Jacques}
\email{vincent.jacques@umontpellier.fr}
\affiliation{Laboratoire Charles Coulomb, Universit\'e de Montpellier and CNRS, 34095 Montpellier, France}

\begin{abstract}
The negatively-charged boron vacancy (\VB) center in hexagonal boron nitride (hBN) is currently garnering considerable attention for the design of two-dimensional (2D) quantum sensing units. Such developments require a precise understanding of the spin-dependent optical response of \VB centers, which still remains poorly documented despite its key role for sensing applications. Here we investigate the spin-dependent photodynamics of \VB centers in hBN by a series of time-resolved photoluminescence (PL) measurements. We first introduce a robust all-optical method to infer the spin-dependent lifetime of the excited states and the electron spin polarization of \VB centers under optical pumping. Using these results, we then analyze PL time traces recorded at different optical excitation powers with a seven-level model of the \VB center and we extract all the rates involved in the spin-dependent optical cycles, both under ambient conditions and at liquid helium temperature. These findings are finally used to study the impact of a vector magnetic field on the optical response. More precisely, we analyze PL quenching effects resulting from electron spin mixing induced by the magnetic field component perpendicular to the \VB quantization axis. All experimental results are well reproduced by the seven-level model, illustrating its robustness to describe the spin-dependent photodymanics of \VB centers. This work provides important insights into the properties of \VB centers in hBN, which are valuable for future developments of 2D quantum sensing units.
\end{abstract} 

\maketitle

\section{Introduction} 

During the past decade, quantum sensors based on optically-active spin defects in semiconductors have found a broad variety of applications, in both basic and applied science, due to their unprecedented combination of sensitivity, spatial resolution and ability to operate under a wide range of experimental conditions~\cite{RevModPhys.89.035002}. While the most prominent example is undoubtedly the nitrogen-vacancy (NV) center in diamond~\cite{DohertyReview,rondin_magnetometry_2014,casola_probing_2018,Aslam2023,Finco2023}, the exploration of alternative spin defects and host materials remains an active field of research worldwide~\cite{Awschalom2018,Higginbottom2022,Castelletto_2024,Luo2024,doi:10.1080/23746149.2023.2206049}. In this context, the negatively-charged boron vacancy (\VB) center in hexagonal boron nitride (hBN) is currently attracting a growing interest for the development of quantum sensing and imaging technologies on a two-dimensional material platform~\cite{Gottscholl2020,doi:10.1080/23746149.2023.2206049}. This point defect, which can be readily created by various irradiation methods~\cite{IgorImplant2020,ACSomega2022,LaserWriting2021,nano11061373,Li2021}, has a spin triplet ground level whose electron spin resonance frequencies can be measured optically~\cite{Gottscholl2020} and strongly depends on external perturbations. Despite a very low quantum efficiency of its optical transition, dense ensemble of \VB centers in hBN flakes can be employed as local sensors of static magnetic fields, strain, electric fields, temperature and magnetic noise~\cite{GottschollNatCom2021,ACSPhot_Guo2021,GaoStrain2022,IgorStrain2022,Udvarhelyi2023,PhysRevApplied.18.L061002,Du2021,Tetienne2023,Li2023,Robertson2023}. Compared to other quantum sensing platforms, one advantage of the hBN host material is its high flexibility for device integration. In addition,  \VB centers can be embedded in few-atomic-layer thick hBN flakes without impairing their magneto-optical properties~\cite{DurandPRL2023}, opening appealing perspectives for quantum sensing and imaging with atomic-scale proximity to the sample being probed. \\
\indent Despite rapid developments of hBN-based quantum sensing units during the last years~\cite{doi:10.1080/23746149.2023.2206049}, the spin-dependent photodymanics of \VB centers remains poorly documented~\cite{ivady2020,Reimers2020,baber2021excited,Whitefield2023}. In particular, the understanding of intersystem crossing (ISC) transitions to metastable levels, which are responsible for electronic spin polarization and readout, is still incomplete. In this work, we study the spin-dependent optical response of \VB centers by a series of experiments including measurements of (i) the spin-dependent lifetime of the excited states, (ii) the electron spin polarization under optical pumping, and (iii) spin-dependent photoluminescence (PL) time traces. By analyzing experimental results with a seven-level model of the \VB centers, we extract all the rates involved in the spin-dependent optical cycles, both under ambient conditions and at liquid helium temperature. We then use these findings to analyze the impact of a vector magnetic field on the optical properties of \VB centers in hBN. Similar to the NV defect in diamond~\cite{Tetienne_2012}, we observe a reduced PL intensity when the magnetic field component perpendicular to the \VB quantization axis increases. These results, which are well reproduced by the seven-level model without any fitting parameter, might find applications for all-optical magnetic imaging with \VB centers.

\section{Methods}
\subsection{Energy level structure of the \VB center}

\begin{figure}[t!]
  \centering
  \includegraphics[width = 8.6cm]{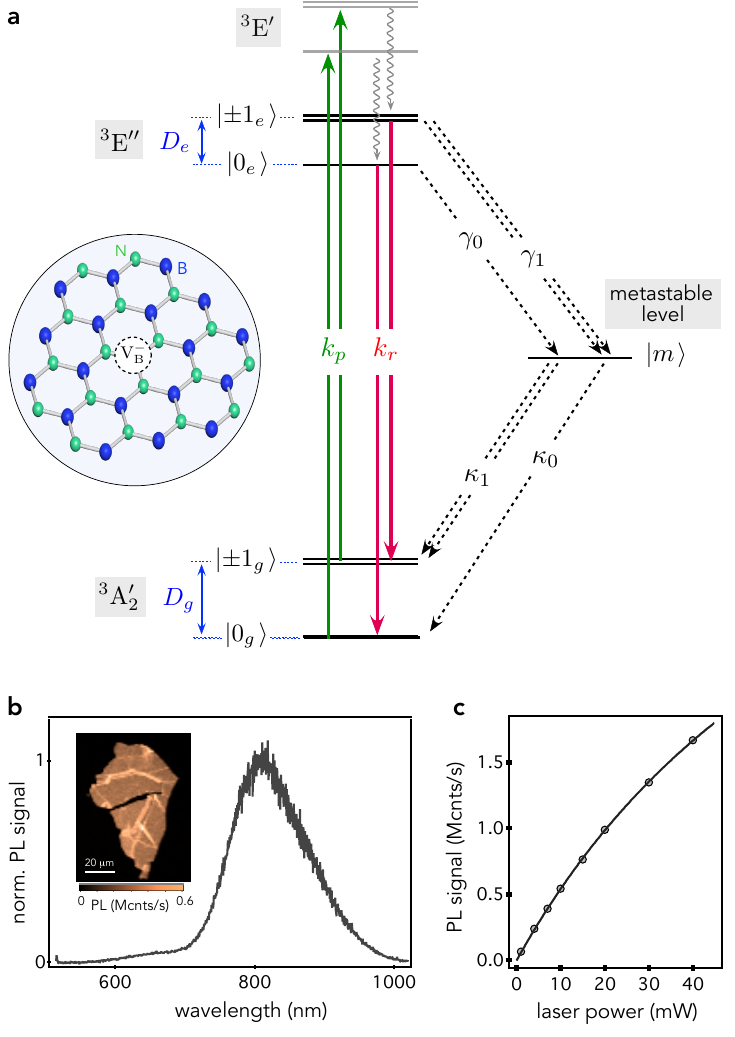}
  \caption{(a) Simplified energy level structure of the \VB center in hBN. The notations used to label the ground ($^{3}{\rm A}_2^{\prime}$) and excited ($^{3}{\rm E}^{\prime\prime}$,$^{3}{\rm E}^{\prime}$) triplet levels refer to the irreducible representations of $\mathcal{D}_{3h}$ symmetry~\cite{ivady2020}. All other notations are defined in the main text. (b) PL spectrum of \VB centers recorded under green laser illumination at room temperature. Inset: Typical PL raster scan of an irradiated hBN flake recorded with a green laser power of $4$~mW. (c) PL signal as function of the laser intensity. The solid line is a fit with a saturation function, yielding a saturation power of the optical transition $\mathcal{P}_{s}=89(3)$~mW.}
  \label{fig1}
\end{figure}

\indent The energy level diagram used to model the spin-dependent optical response of the \VB center is sketched in Fig.~\ref{fig1}(a). The ground level is a spin triplet $^{3}{\rm A}_2^{\prime}$ with an axial zero-field splitting $D_{g}\sim3.47$~GHz between a singlet state with spin projection $m_s = 0$ and a doublet $m_s = \pm1$~\cite{Gottscholl2020}, which are labelled $\left|0_g \right.\rangle$ and $\left|\pm 1_g \right.\rangle$, respectively. The notation $m_s$ refers to the electron spin projection along the quantization axis of the \VB center, that corresponds to the c-axis of the hBN crystal. 

Under optical illumination with a green laser, the \VB center can be promoted to a spin triplet excited level $^{3}{\rm E}^{\prime}$ through a dipole-allowed transition~\cite{ivady2020,Reimers2020}, and then relaxes rapidly by internal conversion towards a lower-energy triplet level $^{3}{\rm E}^{\prime\prime}$ featuring an axial zero-field splitting $D_{e}\sim2.1$~GHz between the $m_s = 0$ ($\left|0_e \right.\rangle$) and $m_s = \pm1$ ($\left|\pm 1_e \right.\rangle$) spin projections~\cite{baber2021excited,yu2021excitedstate,mathur2021excitedstate,mu2021excitedstate}.  Phonon-assisted relaxation from the $^{3}{\rm E}^{\prime\prime}$ excited level to the $^{3}{\rm A}_2^{\prime}$ ground level leads to the emission of a broadband PL signal in the near infrared with a rate $k_r$~\cite{Libbi2022}. Importantly, this emission is intrinsically weak because optical transitions between the $^{3}{\rm E}^{\prime\prime}$ and $^{3}{\rm A}_2^{\prime}$ levels are forbidden by symmetry. Theoretical studies estimated a radiative decay rate in the range of $k_r\sim 10^5$~s$^{-1}$~\cite{ivady2020,Reimers2020}. In the following, we consider that the absorption and the phonon-assisted radiative relaxation channels are purely spin conserving ($\Delta m_s=0$) with identical rates for all electron spin projections [Fig.~\ref{fig1}(a)].

Another relaxation pathway from the $^{3}{\rm E}^{\prime\prime}$ excited level involves non-radiative intersystem crossing (ISC) to a manifold of singlet levels~\cite{ivady2020,Reimers2020}, which is modeled by a single metastable level labelled $\left|\rm{m} \right.\rangle$. 
These processes are spin dependent, giving rise to (i)  polarization of the \VB center in the $\left|0_g \right.\rangle$ ground state by optical pumping and (ii) spin-dependent PL emission that enables optical detection of magnetic resonances~\cite{Gottscholl2020}. As depicted in Fig.~1(a), we denote by $\gamma_0$ (resp. $\gamma_1$) the ISC rate from the excited state $\left|0_e \right.\rangle$ (resp. $\left|\pm 1_e \right.\rangle$) to the metastable level, and by $\kappa_0$ (resp. $\kappa_1$) that from the metastable level to the ground state $\left|0_g \right.\rangle$ (resp. $\left|\pm1_g \right.\rangle$). 

Rate equations within the seven-level model sketched in Fig.~\ref{fig1}(a) are used to analyze the spin-dependent photodynamics of \VB centers.

\subsection{Experimental details}
\indent We investigate ensembles of \VB centers created in few tens of nanometers thick hBN flakes by nitrogen ion irradiation at $30$~keV energy with a dose of $10^{14}$~ions/cm$^2$. The flakes were obtained by mechanical exfoliation of an isotopically-purified h$^{11}$B$^{15}$N crystal produced through metal flux growth methods~\cite{EdgarAdvMat}. 
  \begin{figure*}[t]
  \centering
  \includegraphics[width = 17cm]{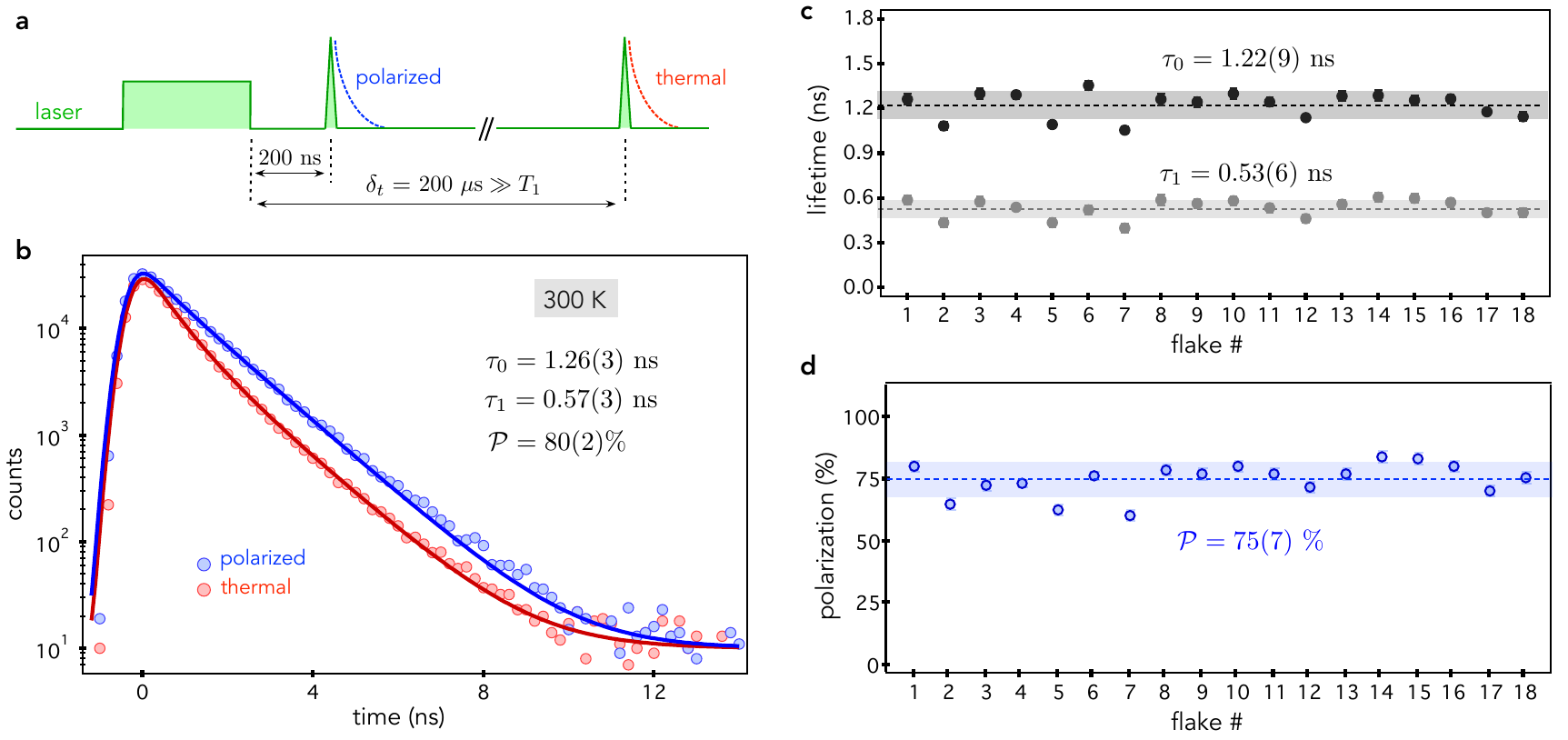}
  \caption{(a) All-optical experimental sequence used to infer the spin-dependent lifetime of the excited states $\{\tau_0,\tau_1\}$ and the electron spin polarization $\mathcal{P}$ of \VB centers at room temperature. The power of the $5$-$\mu$s-long laser pulse is $4$~mW. (b) Time-resolved PL decay recorded within the same experimental sequence for \VB centers prepared in polarized (blue) and thermal (red) spin distributions at room temperature. The PL decays are not normalized and superimposed for clarity using a time offset. The solid lines are data fitting with a bi-exponential function following the procedure described in the main text, leading to $\tau_0=1.26(3)$~ns, $\tau_1=0.57(3)$~ns and $\mathcal{P}=80(2)\%$. (c) Spin-dependent lifetimes and (d) electron spin polarization measured for 18 hBN flakes. The dashed lines indicate the average values of $\{\tau_0,\tau_1,\mathcal{P}\}$ and the shaded areas illustrate the one standard deviation uncertainty of the statistical distribution. All measurements are performed at zero external magnetic field.}
  \label{fig2}
\end{figure*}

The spin-dependent optical response of \VB centers is first studied at room temperature using a scanning confocal microscope equipped with a $0.42$ numerical aperture objective. Optical excitation at $532$~nm is provided by a continuous laser source combined with an acousto-optic modulator to produce $\mu$s-long laser pulses. The resulting diffraction-limited laser spot on the sample surface is on the order of $\sim 1 \ \mu{\rm m}^2$. To measure the excited-state lifetime, we add a pulsed laser source at the same wavelength with a pulse duration of $60$~ps. The PL signal produced by \VB centers in the near infrared is detected by an avalanche photodiode operating in the single photon counting regime. Time-resolved PL measurements are obtained by recording photon detection events on a time tagger with a bin width of $200$~ps. All measurements performed at room temperature are done at zero external magnetic field. The same laser sources and detection modules are also used to carry out experiments at liquid helium temperature ($T=4$~K) with a second scanning confocal microscope, which employs a higher numerical aperture objective  (${\rm NA}=0.82$) and is equipped with a vectorial magnet.  \\
\indent A typical PL raster scan of an irradiated hBN flake recorded at room temperature is shown in the inset of Fig. 1(b). It reveals a uniform PL signal featuring the characteristic broadband emission spectrum of \VB centers. For all the experiments reported in this work, the laser excitation power is below the saturation of the optical transition [Fig. 1(c)]. The longitudinal spin relaxation time of \VB centers is $T_1\sim 13 \ \mu$s at room temperature, and reaches few milliseconds at liquid helium temperature (see Appendix A). 

\section{Results}
\subsection{Spin-dependent excited state lifetimes and electron spin polarization measurements}
\label{SpinExP}

\indent We start by introducing a simple all-optical method to infer both the spin-dependent lifetime of the excited states and the electron spin polarization of \VB centers. The experimental sequence is sketched in Fig.~\ref{fig2}(a). A $5$-$\mu$s-long laser pulse is first used to polarize the \VB centers in the ground state $\left|0_g \right.\rangle$ by optical pumping. After a time delay $\delta_p=200$~ns ensuring that all populations of the metastable level have relaxed to the ground level (see Section~\ref{LifeMeta}), we record the time-resolved PL signal resulting from a $60$-ps laser pulse excitation. Since $\delta_p$ is much shorter than the longitudinal spin relaxation time $T_1$, this measurement is used to analyse the PL decay for spin polarized \VB centers. Within the same experimental sequence, a second ps laser pulse is applied after a time delay $\delta_t=200 \ \mu$s, which is much longer than $T_1$ at room temperature. In this case, the recorded PL decay is thus associated to \VB centers prepared in a thermal spin distribution, {\it i.e.} with equal initial populations in the three spin projections of the ground level. Typical time-resolved PL signals recorded for \VB centers in polarized and thermal spin distributions at room temperature are shown in Fig.~\ref{fig2}(b).

Owing to spin-dependent ISC to the metastable level, the PL signal follows a bi-exponential decay of the form $\mathcal{S}(t)\propto n_{0_e} e^{-t/\tau_0}+n_{\pm 1_e}e^{-t/\tau_1}$, where $n_{0_e}$ and $n_{\pm 1_e}$ are the populations promoted in the excited states $\left|0_e \right.\rangle$ and $\left|\pm1_e \right.\rangle$ by the laser pulse excitation, while $\tau_0$ and $\tau_1$ denote the corresponding spin-dependent lifetimes. Assuming that optical excitation is spin conserving with an identical oscillator strength for the different spin projections, the PL decay can be expressed as 
\begin{equation}
\label{EqDecay}
\mathcal{S}(t)\propto n_{0_g} e^{-t/\tau_0}+n_{\pm 1_g}e^{-t/\tau_1} \ ,
\end{equation} 
where $n_{0_g}$ and $n_{\pm 1_g}$ are the populations in the ground states $\left|0_g \right.\rangle$ and $\left|\pm1_g \right.\rangle$ before optical excitation, which fulfill the relation $n_{0_g}+n_{\pm 1_g}=1$. The experimental data are fitted with this bi-exponential function convolved with the independently-measured instrument response function. 

We first fit the PL decay recorded for \VB centers prepared in a thermal spin distribution [red curve in Fig.~\ref{fig2}(b)]. In this case, the ground state spin populations are fixed to $n_{0_g}=1/3$ and $n_{\pm 1_g}=2/3$, while $\tau_0$ and $\tau_1$ are used as fitting parameters. We obtain $\tau_0=1.26(3)$~ns and $\tau_1=0.57(3)$~ns. These values are then held fixed to fit the PL decay recorded for polarized \VB centers using solely $n_{0_g}$ as a free parameter [blue curve in Fig.~\ref{fig2}(b)], leading to $n_{0_g}=80(2) \%$, which corresponds to the electron spin polarization $\mathcal{P}$ in state $\left|0_g \right.\rangle$. We note that within the fitting uncertainty, the absolute maximum of the bi-exponential decay is identical for the polarized and thermal spin distributions, supporting our assumption that the oscillator strength of the optical transition is identical for all electron spin projections. The same measurements were carried out for \VB centers hosted in 18 different hBN flakes [Fig.~\ref{fig2}(c,d)]. A statistical analysis of the results leads to $\tau_0=1.22(9)$~ns, $\tau_1=0.53(6)$~ns and $\mathcal{P}=75(7) \%$, where the uncertainties correspond to one standard deviation of the statistical distribution.

We now link these measurements to transition rates in the seven-level model of the \VB center sketched in Fig.~1(a). On the one hand, the spin-dependent lifetimes are simply given by 
\begin{equation}
\label{Cons1}
\tau_0=(k_r+\gamma_0)^{-1} \ {\rm and} \ \tau_1=(k_r+\gamma_1)^{-1} \ .
\end{equation} 
On the other hand, it can be shown by rate equations that the steady state populations in the ground states $\{n^s_{0_g},n^s_{\pm1_g}\}$ after optical pumping with the $5$-$\mu$s-long laser pulse fulfill the relation (see Appendix B)
\begin{equation}
\label{EqP}
\frac{n^s_{0_g}}{n^s_{\pm1_g}}=\frac{\mathcal{P}}{1-\mathcal{P}}=\frac{\kappa_0}{2\kappa_1}\times\frac{\gamma_1}{\gamma_0}\times\frac{k_{r}+\gamma_0}{k_{r}+\gamma_1} \ .
\end{equation}
 
Considering that $k_r\sim 10^5$~s$^{-1}$~\cite{ivady2020,Reimers2020}, the radiative decay rate can be safely neglected such that 
\begin{equation}
\label{fix}
 \gamma_{0}=\tau_{0}^{-1} \ , \ \gamma_{1}=\tau_{1}^{-1} \ , \ {\rm and} \ \ \frac{\kappa_0}{\kappa_1}=\frac{2\mathcal{P}}{1-\mathcal{P}} \ .
\end{equation}

\subsection{Lifetime of the metastable level}
\label{LifeMeta}

\begin{figure}[t!]
  \centering
  \includegraphics[width = 8.6cm]{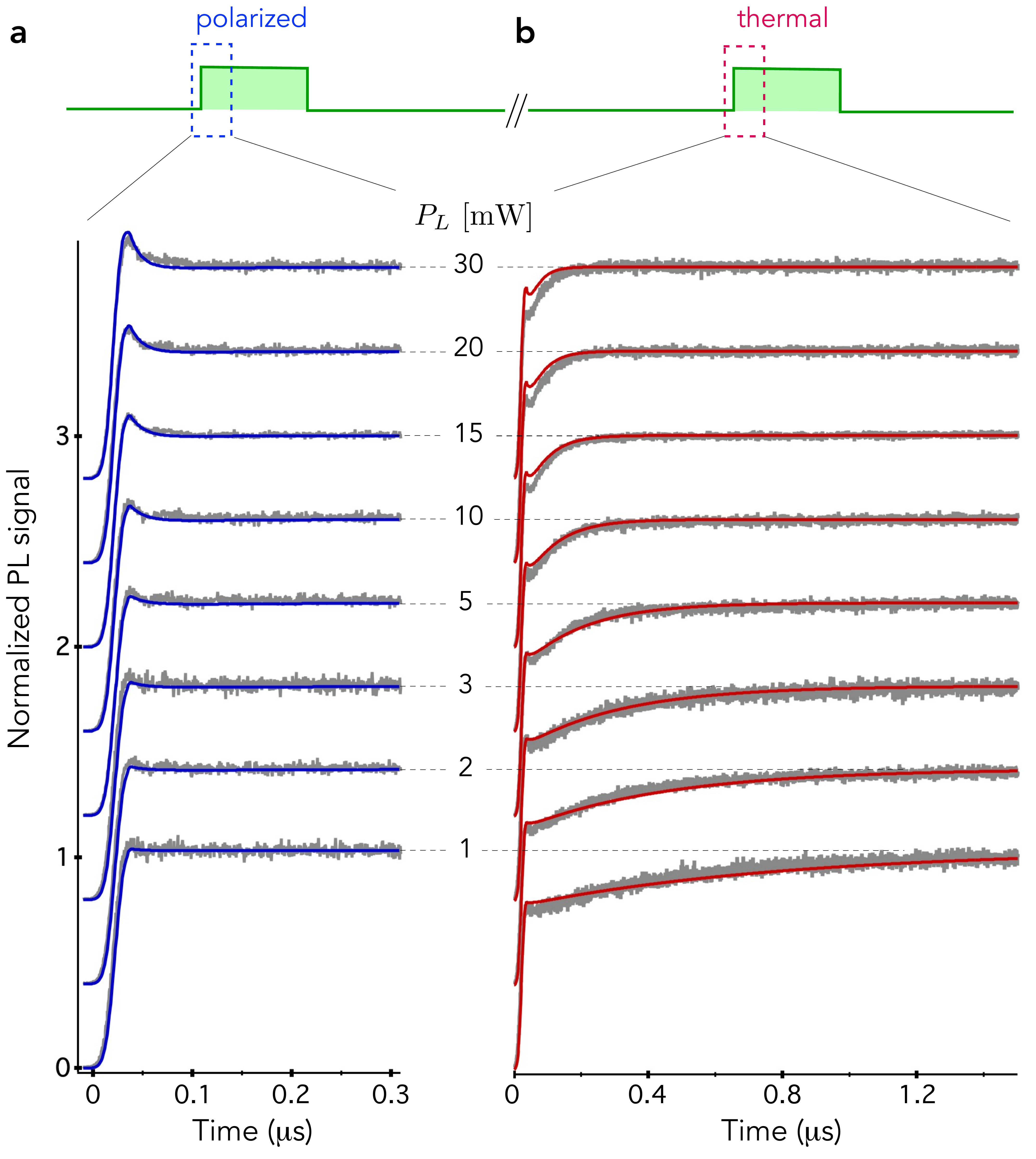}
  \caption{Normalized PL time traces recorded at different laser powers for \VB centers prepared in (a) polarized and (b) thermal spin distributions at room temperature. The data are vertically shifted for clarity and the solid lines are the results of a global fit of the dataset using the seven-level model of the \VB center.}
  \label{fig3}
\end{figure}

\indent These first measurements are then used to estimate the lifetime of the metastable level at room temperature. To this end, we record PL time traces produced by a 5-$\mu$s-long laser pulse with varying power for \VB centers prepared in polarized and thermal spin distributions. The results are summarized in Fig.~\ref{fig3}. For polarized \VB centers, a sharp PL peak appears at the beginning of the pulse when the laser power increases [Fig.~\ref{fig3}(a)]. This effect, which is commonly observed for other optically-active spin defects such as the NV center in diamond~\cite{PhysRevB.74.104303,Robledo2011}, results from ISC to the metastable level. For \VB centers prepared in a thermal state, the PL signal starts at a lower value owing to spin-dependent emission, and then increases before reaching a steady-state corresponding to the PL signal obtained for polarized \VB centers [Fig.~\ref{fig3}(b)]. 

The full dataset is simultaneously fitted with the seven-level model, while constraining the values of $\gamma_0$, $\gamma_1$ and $\kappa_0/\kappa_1$ by the measurements described in the previous section [see Eqs.~(\ref{fix})]. Optical absorption is described by a rate $k_p=k_p^0\times P_L$, where $P_L$ is the laser power focused on a diffraction-limited spot around $1 \ \mu{\rm m}^2$. The fit then uses only $k_p^0$ and the ISC rate $\kappa_0$ as free parameters. Note that the temporal profile of the laser pulse was independently measured by detecting the light reflected by the sample. This profile is integrated into the fitting procedure to take into account the rising edge of the laser pulse. Data fitting leads to $k_p^0=1.0(4)\times 10^6$~s$^{-1}$/mW and $\kappa_0=41(8)\times 10^6$~s$^{-1}$, where the uncertainties are evaluated by taking into account those on the parameters $\gamma_0$, $\gamma_1$ and $\kappa_0/\kappa_1$. The value obtained for $k_p^0$ is in good agreement with a recent report~\cite{DurandPRL2023} and the lifetime of the metastable level is given by $\tau_{\rm m}=(\kappa_0+2\kappa_1)^{-1}$, leading to $\tau_{\rm m}=18(3)$~ns. This very short metastable lifetime explains why the PL signal of \VB centers can be detected despite the very low quantum efficiency of the optical transition. Given the rising and falling edges of the laser pulse, it was not possible to directly measure the metastable level lifetime through PL recovery methods~\cite{PhysRevB.74.104303,Robledo2011}. The transition rates of the seven-level model inferred at room temperature are summarized in Table~1.\\
\indent Even though the fitting procedure reproduces fairly well the full experimental dataset, the model slightly overestimates the initial PL signal for \VB centers in a thermal state, with a deviation being more pronounced at the highest laser powers [Fig.~\ref{fig3}(b)]. It has recently been suggested that the direct relaxation from the $^{3}{\rm E}^{\prime\prime}$ excited level to the $^{3}{\rm A}_2^{\prime}$ ground level might involve an additional non-radiative component with a rate $k_{nr}$~\cite{Whitefield2023}, which competes with ISC transitions to the metastable level. To check this hypothesis, the fit was performed by introducing a direct, non-radiative decay rate as an additional free parameter. In this case $k_p^0$, $\kappa_0$ and $k_{nr}$ are used as fitting parameters of the seven-level model, all the other rates being constrained by Eqs.~(\ref{Cons1}) and~(\ref{EqP}), in which $k_r$ is replaced by $k_{nr}$. As shown in the Appendix~C, we do not obtain a clear improvement of the fit. It is thus not possible to discriminate between the two models with this set of measurements. The fit leads to a strong non-radiative decay rate $k_{nr}=8(1)\times 10^{8}$~s$^{-1}$ and an increased optical pumping rate $k_p^0=10(2)\times 10^6$~s$^{-1}$/mW. Although the ISC transitions rates and their ratio are drastically modified when a direct non-radiative decay is introduced in the model, we obtain a similar lifetime of the metastable level (see Appendix C). \\
\indent In what follows, we keep the simplest model of the \VB center which does not introduce a direct non-radiative decay from the excited level. The discrepancy between the fit and the experiments for the thermal state at short time scale could be caused by charge state conversion of \VB centers under strong optical pumping.

\subsection{Low temperature measurements}
  \begin{figure}[t]
  \centering
  \includegraphics[width = 8.6cm]{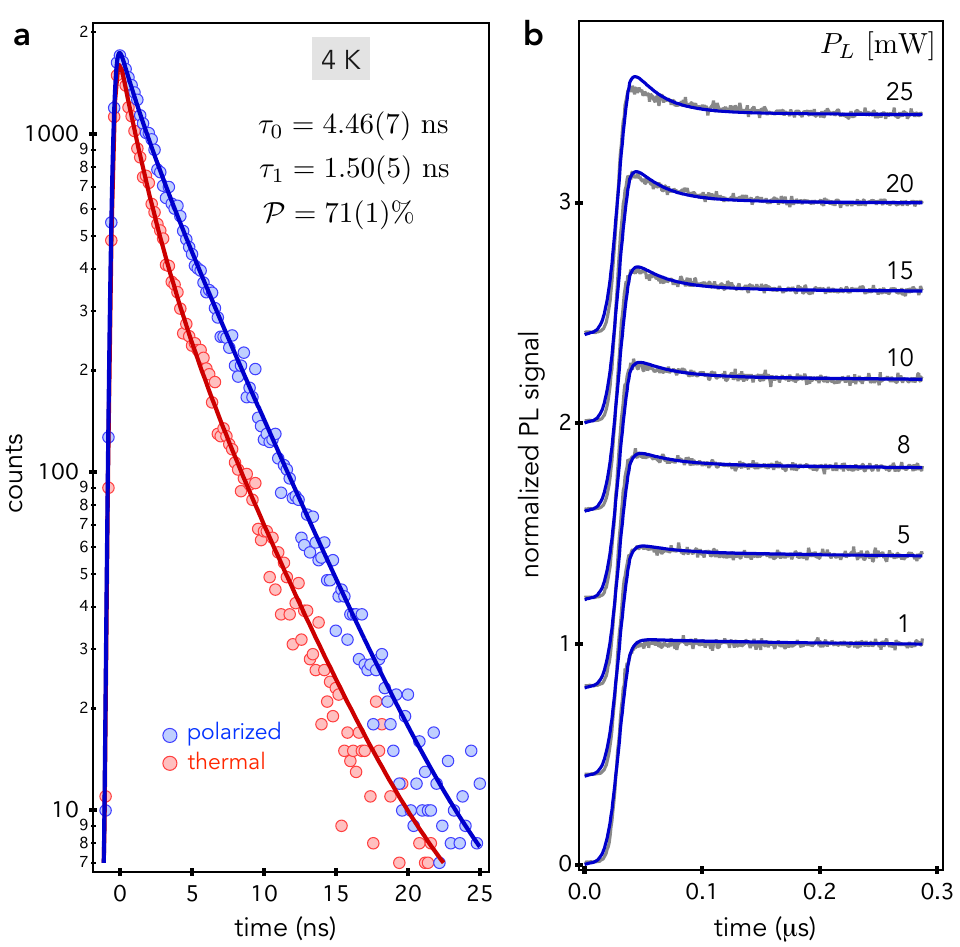}
  \caption{(a) Time-resolved PL decay recorded for \VB centers prepared in polarized (blue) and thermal (red) spin distributions at liquid helium temperature using the experimental sequence sketched in Fig.~\ref{fig2}(a) with $\delta_t=20$~ms. The solid lines are data fitting with a bi-exponential function following the procedure described in section~\ref{SpinExP}, from which we extract $\tau_0$, $\tau_1$ and $\mathcal{P}$. (b) Normalized PL time traces recorded at different laser powers for \VB centers prepared in a polarized spin distribution. The data are vertically shifted for clarity and the solid lines are the results of a global fit of the dataset using the seven-level model of the \VB center, in which the values of the rates $\gamma_0$, $\gamma_1$ and $\kappa_0/\kappa_1$ are constrained by Eqs.~(\ref{fix}).}
  \label{fig4}
\end{figure}

We now investigate how the transition rates of the seven-level model evolve at liquid helium temperature ($T=4$K). The spin-dependent lifetime of the excited states and the electron spin polarization are first measured at $4$K using the method introduced in Section~\ref{SpinExP}. The experimental sequence is identical to that shown in Fig.~2(a), except that the waiting time before applying the second ps laser pulse used to probe the PL decay of the thermal state is increased to $\delta_t=20$~ms because the longitudinal spin relaxation of \VB centers reaches few ms at $4$K (see Appendix A). Time-resolved PL signals recorded for \VB centers in polarized and thermal spin distributions at low temperature are shown in Fig.~\ref{fig4}(a). Data fitting with a bi-exponential decay following the procedure described in Section~\ref{SpinExP} leads to $\tau_0=4.46(7)$~ns, $\tau_1=1.50(5)$~ns and $\mathcal{P}=71(1)\%$. The lifetime of the excited states increases by almost a factor of three. This increase, which has also been reported by others~\cite{mu2021excitedstate,Mendelson2022}, indicates a significant variation of non-radiative ISC rates with temperature, whose detailed analysis is going beyond the scope of the present work. However, we note that the ratio $\tau_1/\tau_0=\gamma_0/\gamma_1$, which is directly linked to the spin-dependent PL contrast, is almost identical to that obtained at room temperature.  
In addition, the electron spin polarization does not significantly vary at $4$~K, suggesting that the ratio of ISC rates $\kappa_0/\kappa_1$ is not drastically modified [See Eq.~(4)]. A statistical analysis over a large number of hBN flakes was not performed at low temperature because the $20$-ms waiting time required to measure the PL decay of \VB centers in a thermal state leads to very long data acquisition times.

The lifetime of the metastable level is then estimated by analyzing PL time traces recorded at different laser powers for spin polarized \VB centers. As shown in Fig.~\ref{fig4}(b), a PL peak is once again detected at the beginning of the pulse when the laser power increases. Fitting the full dataset with the seven-level model leads to $\kappa_0=24(2)\times 10^6$~s$^{-1}$, such that $\tau_m=30(3)$~ns~at~$4$~K. The lifetime of the metastable level is therefore also slightly increased at low temperature. On the other hand, we obtain a decreased optical pumping rate under green laser illumination, $k_p^0=4.4(3) \times 10^5$~s$^{-1}$/mW, despite a smaller focused laser spot owing the larger NA of the microscope objective. This observation is assigned to a modification of the phonon-assisted absorption efficiency of the $^{3}{\rm A}_2^{\prime}\rightarrow \ ^{3}{\rm E}^{\prime}$ optical transition at $T=4$~K [see Fig.~1(a)], whose maximum has been measured around $470$~nm at room temperature~\cite{IgorImplant2020}.

A summary of all the transitions rates of the seven-level model measured under ambient conditions and at liquid helium temperature is given in Table~\ref{tab1}. We note that these values might slightly change with the density of \VB centers, the quality of the hBN crystal and the method used to create \VB centers. 

\begin{table}[t]
\begin{tabular}{ccc}
\hline \\
\vspace{-0.6cm}\\
 & \hspace{0.5cm} $T=300$~K \hspace{0.5cm} & \hspace{0.5cm} $T=4$~K \hspace{0.5cm} \\
\vspace{-0.3cm}\\
\hline \\
\vspace{-0.5cm}\\
\ \ $\gamma_0$ \ \ & $0.82(6)$~ns$^{-1}$ & $0.22(1)$~ns$^{-1}$  \\
\ \ $\gamma_1$ \ \ & $1.8(2)$~ns$^{-1}$ & $0.67(3)$~ns$^{-1}$  \\ 
\ \ $\kappa_0$ \ \ & $41(8)$~$\mu$s$^{-1}$ & $24(2)$~$\mu$s$^{-1}$  \\
\ \ $\kappa_1$ \ \ & $7(2)$~$\mu$s$^{-1}$ & $5(1)$~$\mu$s$^{-1}$  \\
\ \ $\tau_m$ \ \ & $18(3)$~ns & $30(3)$~ns  \\
\vspace{-0.2cm}\\
\hline \\
\end{tabular}
\caption{Summary of the transitions rates in the seven-level model of the \VB center at room temperature and at $T=4$~K. At room temperature, the uncertainties of the rates result from the statistical analysis of spin-dependent lifetime and spin polarization measurements performed on 18 hBN flakes [Fig.~2(c,d)]. At low temperature, these measurements are only performed for one hBN flake, explaining why the relative uncertainties are smaller.}
\label{tab1}
\end{table}

\subsection{Magnetic-field-dependent optical properties}

We finally investigate how the optical properties of \VB centers evolve with an external magnetic field. All the measurements described below are carried out at liquid helium temperature, with superconducting coils providing a vector magnetic field $\mathbf{B}$. The spin Hamiltonian of the \VB center in the ground ($\hat{H}_{g}$) and excited ($\hat{H}_{e}$) triplet levels is expressed as
\begin{equation}
\label{Ham}
\hat{H}_{g(e)}=D_{g(e)}\hat{S}_z^2+\gamma \mathbf{B} \cdot \mathbf{\hat{S}} \ ,
\end{equation}
where $D_{g(e)}$ is the axial zero-field splitting in the ground or excited level, $\gamma=28$~GHz/T is the electron gyromagnetic ratio, and $z$ corresponds to the c-axis of the hBN crystal. At cryogenic temperature, the axial zero-field splitting parameters increase to $D_{g}\sim3.63$~GHz and $D_{e}\sim2.2$~GHz owing to variations of the hBN lattice parameters with temperature~\cite{GottschollNatCom2021,ACSPhot_Guo2021,mathur2021excitedstate,mu2021excitedstate}. Note that the transverse zero-field splitting is not included in the spin Hamiltonian because it has no impact on the results. 

\begin{figure}[b]
  \centering
  \includegraphics[width = 8.8cm]{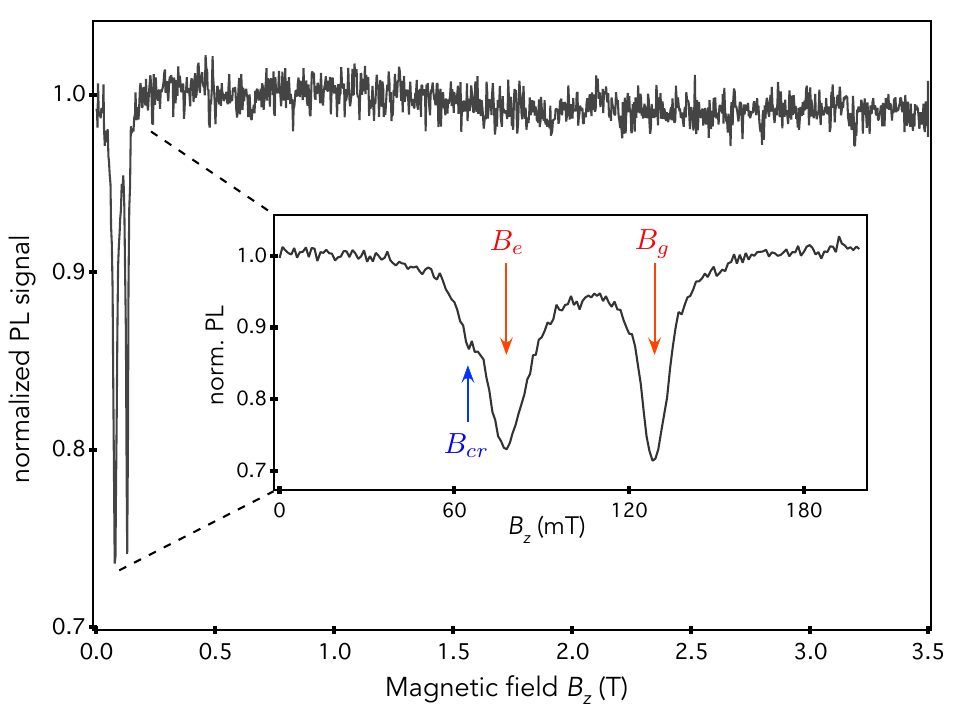}
  \caption{PL intensity of \VB centers as a function of a magnetic field applied along the $z$ axis at cryogenic temperature. The inset shows a zoom on the magnetic field range $0$ - $200$~mT. Anticrossings of the spin states in the ground and excited levels occur at $B_g\sim 129$~mT and $B_e\sim78$~mT, respectively, while cross-relaxation between the \VB center and surrounding $S=1/2$ paramagnetic impurities occurs at $B_{cr}\sim 65$~mT. }
  \label{fig5}
\end{figure}

\begin{figure*}[t]
  \centering
  \includegraphics[width = 18cm]{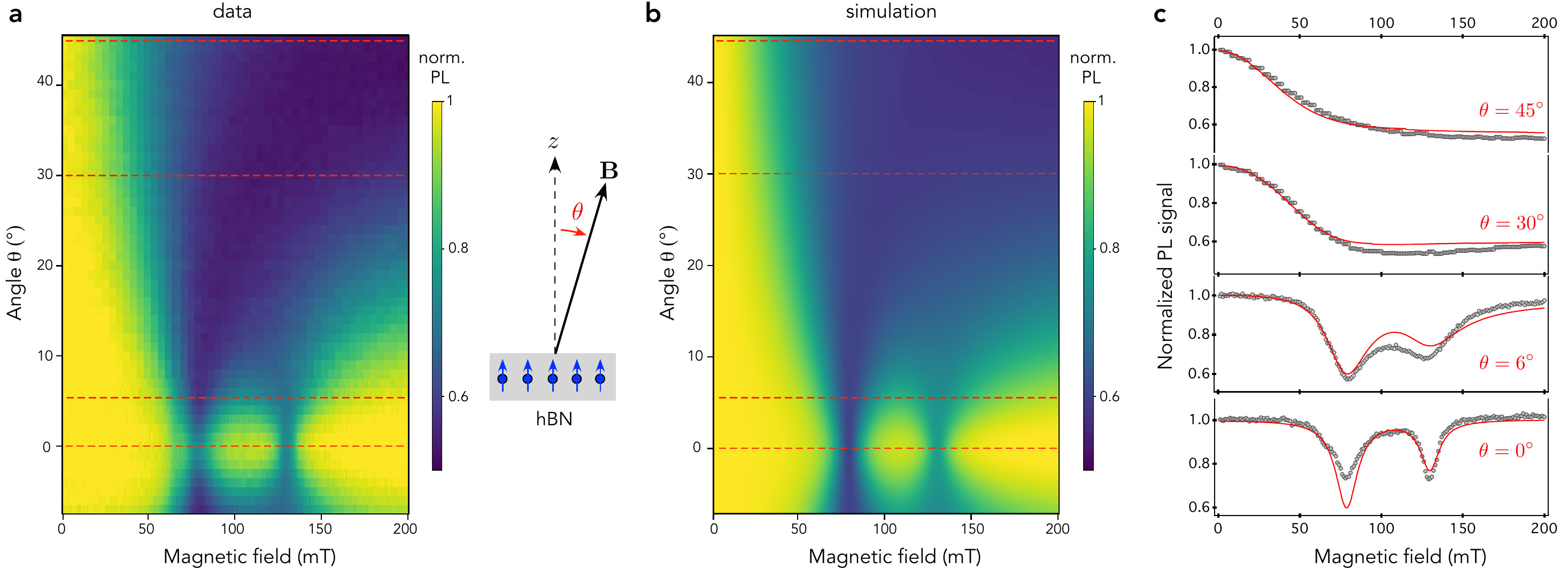}
  \caption{(a) PL intensity of \VB centers as a function of a magnetic field applied with different angles $\theta$ with respect to the $z$ axis. The data are normalized by the PL signal measured at zero field. (b) Magnetic-field-dependent PL intensity calculated through rate equations within the seven-level model of the \VB center. For this calculation we use $D_{g}=3.63$~GHz, $D_{e}=2.2$~GHz, $B^{\rm hf}_{\perp}= 5$~mT, and the transition rates given in Table~1. (c) PL profiles for different angles $\theta$ of the external magnetic field. The black markers show the data and the red solid lines are the result of the calculation.}
  \label{fig6}
\end{figure*}

We start by recording the PL intensity as a function of a magnetic field up to $3.5$~T applied along the $z$ axis. As shown in Fig.~\ref{fig5}, two dips of the PL signal are detected around $B_g\sim D_g/\gamma\sim 129$~mT and $B_e\sim D_e/\gamma\sim 78$~mT, which correspond to anticrossings between the spin states $\left|0_{g(e)} \right.\rangle$ and $\left|-1_{g(e)} \right.\rangle$ in the ground and excited levels, respectively~\cite{baber2021excited,yu2021excitedstate,mathur2021excitedstate,mu2021excitedstate}. Such anticrossings are mediated by the hyperfine coupling of the \VB center with the three neighboring $^{15}$N nuclei [not included in Eq.~(\ref{Ham})]. In a simple picture, the hyperfine interaction can be viewed as an effective classical magnetic field with a component perpendicular to the $z$ axis given by $$B^{\rm hf}_{\perp}\sim \frac{1}{4\gamma}\sum_{i=1}^{3}[\mathcal{A}^{(i)}_{xx}+\mathcal{A}^{(i)}_{yy}] \ ,$$ where $\mathcal{A}^{(i)}_{xx}$ and $\mathcal{A}^{(i)}_{yy}$ denote the transverse components of the hyperfine interaction tensor for each neighboring $^{15}$N nuclear spin. Using ab initio calculations of the hyperfine tensor~\cite{ivady2020}, we estimate that $B^{\rm hf}_{\perp}\sim 5$~mT for \VB centers hosted in a hBN crystal isotopically purified with $^{15}$N~\cite{Clua2023,Gong2024}. At the anticrossings, this effective magnetic field mixes the spin states $\left|0_{g(e)} \right.\rangle$ and $\left|-1_{g(e)}  \right.\rangle$, resulting in a reduced PL signal owing to spin-dependent ISC to the metastable level. The PL dip at the excited level anticrossing exhibits a shoulder at $B_{cr}\sim D_g/2\gamma\sim 65$~mT [Fig.~\ref{fig5}], which corresponds to cross-relaxation between \VB centers and surrounding $S=1/2$ paramagnetic impurities~\cite{baber2021excited,haykal2021decoherence}. Besides these well-identified PL dips, there are no additional features for magnetic fields up to 3.5 T. We can thus conclude that within this field range, there is no evidence of anticrossings between different orbitals of the excited level, as it is for example the case for NV defects in diamond~\cite{PhysRevLett.128.177401}. More generally, this experiment indicates that no additional spin-triplet level with $D<100$~GHz is involved in the spin-dependent photodynamics, thus supporting the simple seven-level model of the \VB center used in this work.

\indent The PL intensity is then measured as a function of the magnetic field while varying its angle $\theta$ with respect to the $z$ axis. The experimental results are summarized in Fig.~\ref{fig6}(a). When the tilt angle $\theta$ increases by few degrees, the transverse component of the external magnetic field $B_{\perp}=B|\sin\theta|$ is added to the effective hyperfine field, leading to a broadening of the PL dips around level anticrossings. For larger angles ($\theta>15^{\circ}$), the transverse magnetic field component $B_{\perp}$ becomes strong enough to effectively mix the electron spin states in the ground and excited levels, even at low field. In this case, the PL signal decreases steadily with the magnetic field before reaching a plateau corresponding to a maximal PL drop of $\sim 50\%$ for which the electron spin states are fully mixed. Such a PL quenching is directly correlated with an overall reduction of the excited level lifetime (see Appendix D)~\cite{Tetienne_2012}.\\
\indent These experimental results are finally compared to a simulation of the magnetic-field-dependent PL response of \VB centers. For each magnetic field, the three eigenstates of the spin Hamiltonian in the ground and excited triplet levels are expressed as linear combinations of the zero-field eigenstates $\{\left|0_{g(e)} \right.\rangle,\left|+1_{g(e)} \right.\rangle,\left|-1_{g(e)} \right.\rangle\}$. We then use rate equations within the seven-level model of the \VB center to calculate the PL signal, following the procedure introduced in Ref.~\cite{Tetienne_2012} for NV centers in diamond. The calculation is performed with the zero-field transition rates measured in the previous section [see Table~1] and $B^{\rm hf}_{\perp}= 5$~mT, which is assumed to be identical in the ground and excited levels. As shown in Figs.~\ref{fig6}(b,c), a good agreement is obtained between the simulation and the experiment, illustrating the robustness of the seven level model for describing the photodynamics of \VB centers in hBN. \\
\indent The magnetic-field dependent PL quenching of \VB centers might find interesting applications for all-optical magnetic imaging~\cite{Tetienne_2012,PhysRevApplied.13.044023}. Quantitative measurements of static magnetic fields are commonly obtained by recording the Zeeman shift of the \VB center's electron spin sublevels through optically-detected magnetic resonance spectroscopy~\cite{PhysRevApplied.18.L061002,Du2021,Tetienne2023}. This method, which requires to combine optical illumination with a microwave excitation, becomes however challenging when the magnetic field exhibits a strong component perpendicular to the \VB quantization axis leading to mixing of the electron spin sublevels. Such a situation will be inevitably reached when nm-thick hBN sensing units will be placed, for instance, on the surface of a ferromagnet where fields in excess of several hundreds of mT are commonly produced. In this case, the magnetic-field dependent PL quenching of \VB centers could be exploited to map regions of the sample producing large stray fields, such as domain walls or skyrmions~\cite{PhysRevMaterials.2.024406}, without the need for microwave excitation.

\section{Conclusion}
To summarize, we have provided a complete picture of the spin-dependent optical response of an ensemble of \VB centers in hBN. By analyzing a series of time-resolved PL measurements with a simplified seven-level model of the \VB center, we inferred all the rates involved in the spin-dependent optical cycles both at room temperature and at $T=4$~K. Our results indicate that the lifetime of the metastable level responsible for electron spin polarization and readout is very short ($\sim20$-$30$~ns), explaining why the PL signal of \VB centers can be detected despite the very low quantum efficiency of its optical transition. The robustness of the seven-level model was then tested by studying the impact of a vector magnetic field on the optical response. We have shown that electron spin mixing induced by the magnetic field component perpendicular to the \VB quantization axis leads to a strong PL quenching, which is well reproduced by the seven-level model without any fitting parameter. This work provides important insights into the properties of \VB centers in hBN, which are valuable for developing quantum sensing and imaging technologies on a two-dimensional material platform. \\

\noindent {\it Acknowledgements} - This work was supported by the French Agence Nationale de la Recherche under the program ESR/EquipEx+ 2DMAG (grant number ANR-21-ESRE-0025) and through the project Qfoil (ANR-23-QUAC-0003), a grant from NanoX in the framework of the ''Programme des Investissements d’Avenir'' (ANR-17-EURE-0009, Q2D-SENS), the program QuantEdu France, and the Institute for Quantum Technologies in Occitanie. Hexagonal boron nitride crystal growth was supported by the Office of Naval Research award N00014-22-1-2582.

\begin{appendix}
\section{Longitudinal spin relaxation measurements}
\begin{figure}[t]
  \centering
  \includegraphics[width = 8.5cm]{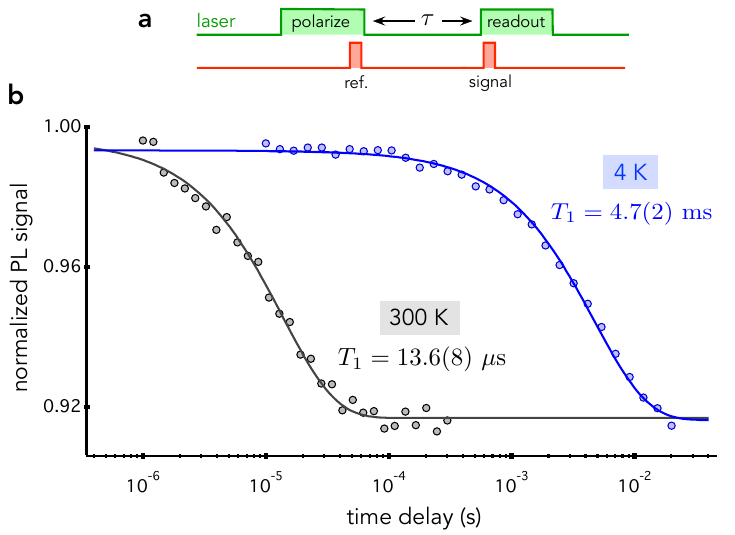}
  \caption{(a) Exprimental sequence used to measure the longitudinal spin relaxation time of \VB centers. The spin-dependent PL signal is integrated at the beginning of the 5-$\mu$s-long readout pulse ($500$~ns window) and normalized using a reference PL value obtained at the end of the first laser pulse. (b) Spin relaxation curves recorded at room temperature (black) and at $T=4$~K (blue). The solid lines are data fitting with an exponential decay.}
  \label{fig1SUP}
\end{figure}

The longitudinal spin relaxation time ($T_1$) of \VB centers was measured with the experimental sequence sketched in Fig.~\ref{fig1SUP}. A 5-$\mu$s-long laser pulse is first used to polarize the \VB centres in the ground state $\left|0_g \right.\rangle$ by optical pumping. After relaxation in the dark during a variable time $\tau$, the remaining population in $\left|0_g \right.\rangle$ is probed by integrating the spin-dependent PL signal produced at the beginning of a second laser pulse. The longitudinal spin relaxation time $T_1$ is then inferred by fitting the decay of the integrated PL signal with an exponential function. 

Typical spin relaxation curves of \VB centers recorded at room temperature and at $T=4$~K are plotted in Fig.~\ref{fig4}(a) in semi-log scale. We obtain $T_1\sim 13.6(8) \ \mu$s at room temperature, a value similar to that usually observed for \VB centres in hBN, which is limited by spin-phonon interactions\cite{Gottscholl2021,Lunghi2022}. At $T=4$~K, the spin relaxation time increases to $T_1\sim 4.7(2)$~ms.

\section{Rate equations in the seven-level model}
Using the seven-level model sketched in Fig.~1(a), the steady state populations of the \VB center in the excited states $\{n^s_{0_e},n^s_{\pm1_e}\}$ after optical pumping with a $5$-$\mu$s-long laser pulse are given by 
\begin{eqnarray}\label{eq:hseq}
n^s_{0_e}&=&\frac{k_p}{k_r+\gamma_0}\times n^s_{0_g}\\
n^s_{\pm1_e}&=&\frac{k_p}{k_r+\gamma_1}\times n^s_{\pm1_g} \ ,
\end{eqnarray}
where $\{n^s_{0_g},n^s_{\pm1_g}\}$ are the steady state populations in the ground states. For the measurements of the spin-dependent lifetime of the excited states, the optical pumping power of the laser pulse used to polarize the \VB centers is such that $k_p\ll(\gamma_0,\gamma_1)$. As a result, $n^s_{0_e}=n^s_{\pm 1_e}\sim 0$. 

On the other hand, the steady state population in the metastable level $n^s_{\rm m}$ fulfill the relations
\begin{eqnarray}\label{eq:hseq}
n^s_{\rm m}&=&\frac{\gamma_1 k_p}{2\kappa_1(k_r+\gamma_1)}\times n^s_{\pm1_g}\\
&=&\frac{\gamma_0 k_p}{\kappa_0(k_r+\gamma_0)}\times n^s_{0_g} \ .
\end{eqnarray}
Since $k_p\ll(\kappa_0,\kappa_1)$ in our measurements, the steady state population in the metastable level can also be neglected. As a result, $n^s_{0_g}+n^s_{\pm1_g}\sim1$ and the electron spin polarization is simply given by $\mathcal{P}\sim n^s_{0_g}$. 

Using Eqs.~(B3) and (B4), we obtain the relationship between $\mathcal{P}$ and the rates of the seven-level model [see Eq.~(3) of the main text]. We finally note that the longitudinal spin relaxation was not included in the model because it has no impact on the spin polarization efficiency as long as the condition $k_p\gg 1/T_1$ is fulfilled.

\section{Data fitting with a direct non-radiative decay rate included in the seven-level model}
\begin{figure}[t]
  \centering
  \includegraphics[width = 8.7cm]{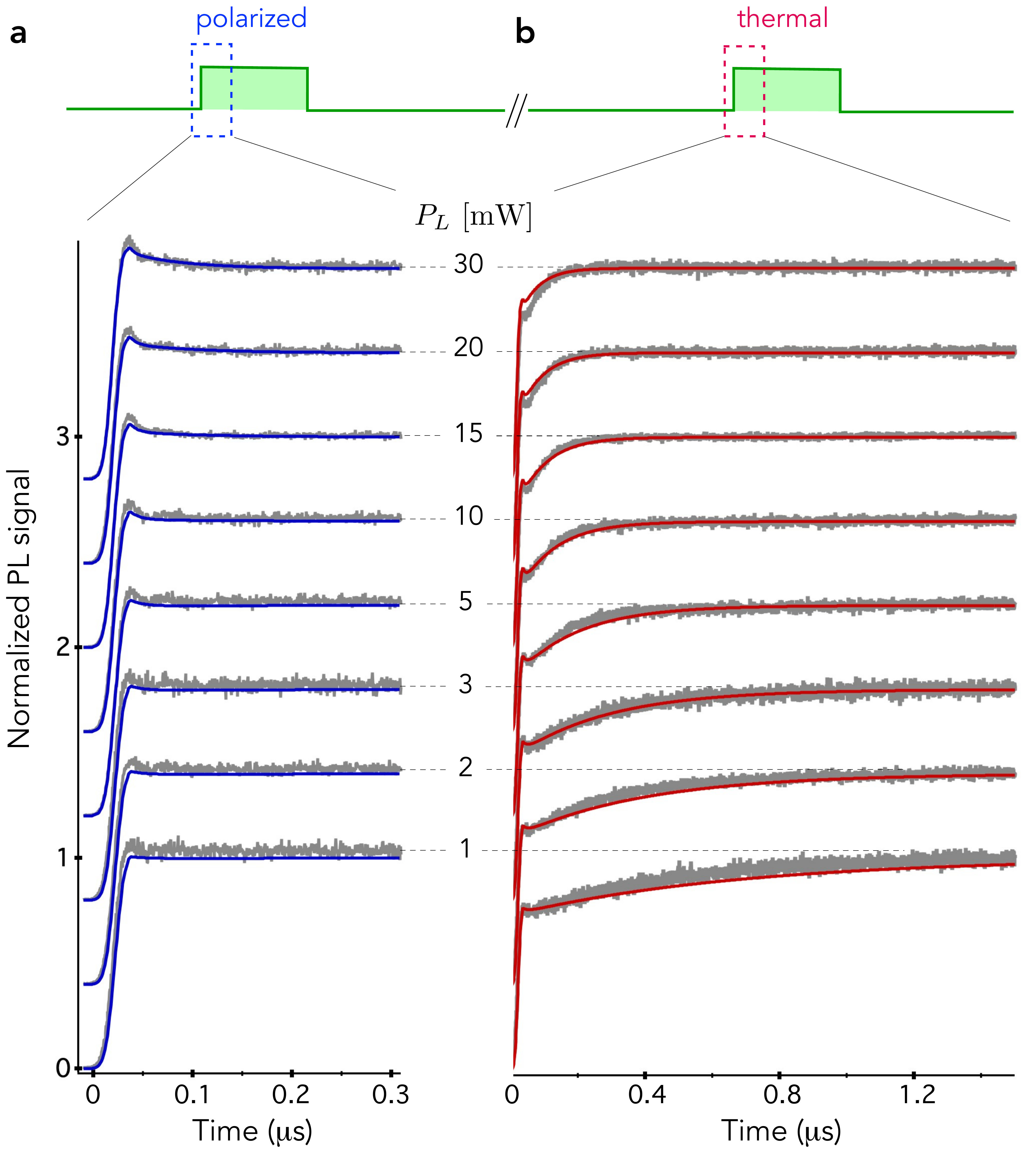}
  \caption{Identical dataset to the one showed in Fig.~3. The solid lines are the results of a global fit of the data with a seven-level model including a direct non-radiative decay rate $k_{nr}$. We use $k_p^0$, $\kappa_0$ and $k_{nr}$ as fitting parameters, all the other rates being constrained by Eqs.~(\ref{Cons1}) and~(\ref{EqP}), in which $k_r$ is replaced by $k_{nr}$.}
  \label{figSUP2}
\end{figure}

The fit of the full dataset of power-dependent PL time traces with a seven-level model including a direct non-radiative decay rate $k_{nr}$ is shown in Fig.~\ref{figSUP2}. We do not obtain a better agreement with the experiments. The model still slightly overestimates the initial PL signal for \VB centers in a thermal state at the highest laser powers. The fit leads to $k_{nr}=8(1)\times 10^{8}$~s$^{-1}$ and a pumping rate $k_p^0=10(2)\times 10^6$~s$^{-1}$/mW. The  ISC rates become $\gamma_0=0.03(2)$~ns$^{-1}$, $\gamma_1=1.1(3)$~ns$^{-1}$, $\kappa_0=9(2)$~$\mu$s$^{-1}$, and $\kappa_1=20(10)$~$\mu$s$^{-1}$, leading to a lifetime of the metastable lifetime $\tau_m=20(10)$~ns.\\

\section{Magnetic field dependent lifetime measurements}
\begin{figure}[h!]
  \centering
  \includegraphics[width = 8.6cm]{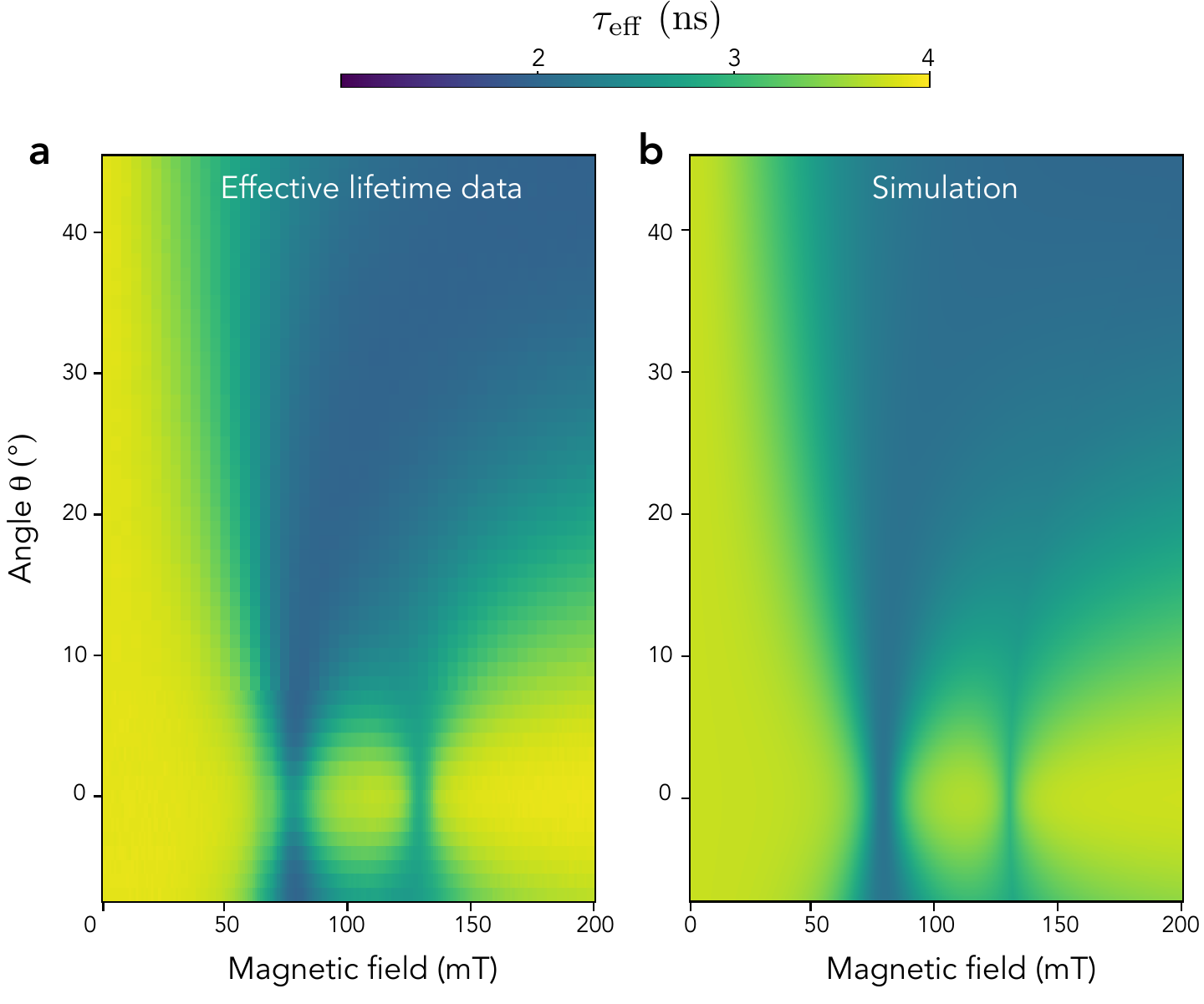}
  \caption{(a) Effective excited-level lifetime $\tau_{\rm eff}$ of \VB centers as a function of a magnetic field applied with different angles $\theta$ with respect to the $z$ axis. (b) Effective lifetime inferred from a calculation using rate equations within the seven-level model of the \VB center. As for the calculation of PL quenching, we use $D_{g}=3.63$~GHz, $D_{e}=2.2$~GHz, $B^{\rm hf}_{\perp}= 5$~mT, and the transition rates given in Table~1.}
  \label{fig3SUP}
\end{figure}

In this appendix, we show that electron spin mixing induced by the magnetic field component perpendicular to the \VB quantization axis leads to an overall reduction of the excited-level lifetime, which is directly correlated with the quenching of the PL signal. To this end, measurements of the PL decay under optical excitation with ps laser pulses are carried out while changing the amplitude and orientation of the external magnetic field. To simplify data analysis, each PL decay is fitted with a single exponential function, from which we infer the {\it effective} lifetime $\tau_{\rm eff}$ of the excited level. The experimental results are summarized in Fig.~\ref{fig3SUP}(a), showing a perfect correlation with the PL quenching map shown in Fig.~\ref{fig6}. Field-induced electron spin mixing increases the mean probability of non radiative ISC transitions to the metastable level, leading to an overall reduction of both the PL signal and the effective lifetime of the excited level.

PL decay curves were simulated with the seven level model, using the same parameters as those used to calculate the magnetic field dependent PL response. The effective lifetime was then obtained by fitting these simulated PL decays with an exponential function. As illustrated by Fig.~\ref{fig3SUP}(b), we obtain once again a very good agreement between the predictions of the seven-level model and the experimental results.

\end{appendix}

%\newpage

%

\end{document}